\documentclass[12pt, preprint]{aastex}
\usepackage{epsfig}
\usepackage{natbib}

\sfcode`A=1000 \sfcode`B=1000 \sfcode`C=1000 \sfcode`D=1000
\sfcode`E=1000 \sfcode`F=1000 \sfcode`G=1000 \sfcode`H=1000
\sfcode`I=1000 \sfcode`J=1000 \sfcode`K=1000 \sfcode`L=1000
\sfcode`M=1000 \sfcode`N=1000 \sfcode`O=1000 \sfcode`P=1000
\sfcode`Q=1000 \sfcode`R=1000 \sfcode`S=1000 \sfcode`T=1000
\sfcode`U=1000 \sfcode`V=1000 \sfcode`W=1000 \sfcode`X=1000
\sfcode`Y=1000 \sfcode`Z=1000

\newcommand{\etal}{{et~al.\/}}

\newcommand{\ie}{{i.e.\/}}
\newcommand{\Lsun}{\hbox{L$_{\odot}$}}
\newcommand{\Msun}{\hbox{M$_{\odot}$}}

\slugcomment{v1 \today}
\shorttitle{}
\shortauthors{Khan \etal}

\begin{document}

\title{On the nature of the first galaxies selected at 350\,$\mu$m}

\author{Sophia A. Khan\altaffilmark{1,2,3,4,5}, 
Pierre F.\ Chanial\altaffilmark{2}, 
S.~P.\ Willner\altaffilmark{4}, 
Chris P.\ Pearson\altaffilmark{6,7}, 
M.~L.~N.\ Ashby\altaffilmark{4},
Dominic J.\ Benford\altaffilmark{3}, 
David L.\ Clements\altaffilmark{2}, 
Simon Dye\altaffilmark{8}, 
Duncan Farrah\altaffilmark{9,10}, 
G.~G.\ Fazio\altaffilmark{4},
J.-S.\ Huang\altaffilmark{4}, 
V.\ Lebouteiller\altaffilmark{10}, 
Emeric Le~Floc'h\altaffilmark{11},
Gabriele Mainetti\altaffilmark{12}, 
S.\ Harvey Moseley\altaffilmark{3}, 
Mattia Negrello\altaffilmark{13},
Stephen Serjeant\altaffilmark{13},
Richard A.\ Shafer\altaffilmark{3},
Johannes Staguhn\altaffilmark{3,14}, 
Timothy J. Sumner\altaffilmark{2}  \and
Mattia Vaccari\altaffilmark{15}
}

\altaffiltext{1}{ALMA Fellow, Pontificia Universidad Cat\'olica,
Departamento Astronom\'ia y Astrof\'isica, 4860 Vicu\~na Mackenna,
Casilla 406, Santiago 22 Chile} 

\altaffiltext{2}{Imperial College London, Blackett Laboratory, Prince
Consort Road, London SW7 2AZ, UK} 

\altaffiltext{3}{Observational Cosmology Laboratory (Code 665), NASA
 Goddard Space Flight Center, Greenbelt, MD 20771 USA} 

\altaffiltext{4}{Harvard-Smithsonian Center for Astrophysics, 60
Garden Street, Cambridge, MA 02138 USA} 

\altaffiltext{5}{Shanghai Key Lab for Astrophysics, Shanghai Normal
University, Shanghai 200234, China} 

\altaffiltext{6}{Rutherford Appleton Laboratory, Chilton, Didcot,
Oxfordshire OX11 0QX, UK} 

\altaffiltext{7}{Department of Physics, University of Lethbridge,
4401 University Drive, Lethbridge, Alberta T1J 1B1, Canada}
 
\altaffiltext{8}{School of Physics and Astronomy, Cardiff University,
Queens Buidings, Cardiff, CF24 3AA, UK} 

\altaffiltext{9}{Astronomy Centre, University of Sussex, Falmer, Brighton UK}

\altaffiltext{10}{Department of Astronomy, Cornell University, 610
Space Sciences Building, Ithaca, NY 14853 USA}

\altaffiltext{11}{{\it Spitzer} Fellow, Institute for Astronomy,
University of Hawaii, 2680 Woodlawn Drive, Honolulu, HI 96815 USA}

\altaffiltext{13}{Department of Physics and Astronomy, Open
University, Walton Hall, Milton Keynes MK7 6AA, UK}

\altaffiltext{14}{Department of Astronomy, University of Maryland,
College Park, MD 20742 USA}

\altaffiltext{15}{Department of Astronomy, University of Padova,
Vicolo Osservatorio 3, I-35122, Padova, Italy}

\begin{abstract}

We present constraints on the nature of the first galaxies selected
at 350\,$\mu$m.  The sample includes galaxies discovered in the
deepest blank-field survey at 350\,$\mu$m (in the Bo\"otes Deep
Field) and also later serendipitous detections in the Lockman Hole.
In determining multiwavelength identifications, the 350\,$\mu$m
position and map resolution of the second generation Submillimeter
High Angular Resolution Camera (SHARC~II) are critical, especially in
the cases where multiple radio sources exist and the 24\,$\mu$m
counterparts are unresolved.  Spectral energy distribution templates
are fit to identified counterparts, and the sample is found to
comprise IR-luminous galaxies at $1<z<3$ predominantly powered by
star formation.  The first spectrum of a 350\,$\mu$m-selected galaxy
provides an additional confirmation, showing prominent dust grain
features typically associated with star-forming galaxies.

Compared to submillimeter galaxies selected at 850 and 1100\,$\mu$m,
galaxies selected at 350\,\micron\ have a similar
range of far-infrared color temperatures.  However, no
350\,$\mu$m-selected sources are reliably detected at 850 or
1100\,$\mu$m.  Galaxies in our sample with redshifts $1<z<2$
show a tight correlation between the far- and
mid-infrared flux densities, but galaxies at higher
redshifts show  a large
dispersion in their mid- to far-infrared colors.  This implies
a limit to which the mid-IR emission traces the far-IR emission in
star-forming galaxies.


The 350\,$\mu$m flux densities ($15<S_{350}<40$\,mJy) place these
objects near the {\it Herschel}/SPIRE 350\,$\mu$m confusion
threshold, with the lower limit on the star formation rate density
suggesting the bulk of the 350\,$\mu$m contribution will come from
less luminous infrared sources and normal galaxies.  Therefore the
nature of the dominant source of the 350\,$\mu$m background ---
star-forming galaxies in the epoch of peak star formation in the
universe --- could be more effectively probed using ground-based
instruments with their angular resolution and sensitivity
offering significant advantages over space-based imaging.

\end{abstract}



\section{Introduction}

Submillimeter-selected galaxies (SMGs) were discovered in pioneering
lensed and blank-field surveys (e.g.,
\citealt{Smail1997, Barger1998, Eales1999}) with the
850\,$\mu$m-optimised Submillimeter 
Common User Bolometer Array (SCUBA; \citealt{Holland1999}) and
later in similar surveys with millimeter detectors (e.g., MAMBO,
BOLOCAM, etc.: \citealt{Bertoldi2000, Laurent2005}).  SMGs
mainly comprise massive, star-forming galaxies (see, e.g.,
\citealt{fox02}) around $z\sim2$ \citep{Chapman2005} with the
bulk of the emission generated at rest-frame far-IR wavelengths.
SMGs are thus part of the IR-luminous galaxy population, which 
includes galaxies
found in the local universe by the Infrared Astronomical
Satellite ({\it IRAS}) All Sky Survey \citep{Soifer1984,
JosephWright1985, Soifer1987} and other galaxies
detected in mid- and far-IR bands, most notably (in
terms of number selected) with the {\it Infrared Space Observatory}
(\citealt{Kessler1996}), the {\it Spitzer Space Telescope}
(\citealt{Werner2004}) and the {\it Akari Infrared Satellite}
(\citealt{Murakami2007}) out to $z\sim2$ (see, e.g.,
\citealt{Rowan-Robinson1997,Puget1999,Aussel1999,Elbaz2002,
Chary2004,Lonsdale2004,LeFloch2004,Yan2004a,LeFloch2005,Matsuhara2006}). 

The majority of SMGs have been selected at long
submillimeter--millimeter bands (500--1300\,$\mu$m).  Shorter
submillimeter wavelengths (200--500\,$\mu$m) are more demanding for
ground-based observers. For example, on a  good night at
Mauna Kea, atmospheric transmission is about  $\sim$30\% at
350\,\micron\ but $\ga$80\%\ 
at 850\,$\mu$m; \citep{Serabyn1998}.  Despite this, the 
first galaxy selected {\it purely} by 350\,$\mu$m
emission --- SMM J143206.65+341613.4 (=SSG~1, Short Submillimeter
Galaxy~1) --- was discovered (\citealt{Khan2005}) in a deep, blank
survey of the Bo\"otes Deep Field with the second generation
Submillimeter High Angular Resolution Camera (SHARC~II;
\citealt{Dowell2003, Moseley2004}) along with a second detection, SMM
J143206.11+341648.4 (=SSG~2; \citealt{Khan2006,Khan2007}).  This
survey, reaching 13\,mJy and
currently the deepest at 350\,\micron, obtained the first
constraints on the 350\,$\mu$m source counts.
Three additional 350\,$\mu$m-selected galaxies have  been found
in serendipitous SHARC~II follow-up observations of 850\,$\mu$m and
1100\,$\mu$m SMGs in the Lockman Hole (\citealt{Laurent2006,
Coppin2008}).

Short-wavelength submillimeter surveys are expected to principally select 
star-forming galaxies at $1<z<3$ (see, e.g., \citealt{PK2009,
Khan2007, Khan2006}), the epoch of peak star formation in the
universe (e.g., \citealt{Hopkins2006}). Source count models that
reproduce the observed 350\,$\mu$m counts include a predominantly IR-luminous
galaxy population evolving with redshift (e.g.,
\citealt{PK2009,Franceschini2009}),
but more observations are needed to verify these predictions.

This paper presents a detailed characterization of each of the five
350\,$\mu$m-selected galaxies following the approach of
\citet{Khan2005}.  Multiwavelength data are presented, and spectral
energy distribution (SED) template fitting is used to provide
constraints on the photometric redshifts, thermal parameters (IR
luminosity and dust temperature), and energy diagnostics.
Additionally the first mid-IR spectrum of a 350\,$\mu$m-selected
galaxy, obtained using {\it Spitzer}'s Infrared Spectrograph
(\citealt{Houck2004}), is given.  The properties of the sample are
compared with 850 and 1100\,$\mu$m-selected SMGs.  Table~\ref{tab:lh}
identifies the five sources and includes short nicknames used for
convenience.  The {\it WMAP} first year cosmological parameters ($\rm
H_0=71\,km\,s^{-1}\, Mpc^{-1}$, $\rm \Omega_m$=0.27, $\rm
\Omega_\Lambda$=0.73; Bennett et al. 2003) are used throughout this
work.

\section{Observations, data reduction, and counterpart identification}

\subsection{Observations}

The SHARC~II observations, data reduction, and source extraction
procedure for the Bo\"otes Deep Field survey are discussed in detail by
\citet{Khan2007}. Further details of source extraction are given
by \citet{Khan2006}.  Monte Carlo simulations for determining the
survey completeness also provide a measure of the flux boosting.
Approximately 4000 artificial sources of random intensity and
position were inserted into the raw data map, then extracted using
the same procedure as for the real sources (\citealt{Khan2007}).  For
recovered $\geq$3$\sigma$ sources with input flux densities
$>$15\,mJy, the ratio of measured flux density to input flux density
indicates the average flux boosting is 1.07$\pm$0.23.  Given the
multiwavelength confirmation on SSG~1 and SSG~2, low outliers can be
ruled out.  Therefore we proceed without applying a correction factor
to these sources and assume this also holds for the sources in the
Lockman Hole.  For the Lockman Hole, published 350\,$\mu$m flux
densities were combined with public $B$, $R$, $I$, and $z$ photometry
from the Subaru Lockman Hole survey (\citealt{Dye2008}), $K$ band imaging from UKIDSS  
(\citealt{Lawrence2007}), archival {\it Spitzer} IRAC and MIPS
imaging, and 1.4\,GHz data from \citet{Biggs2006}.  These
compiled photometric data are presented in Table~\ref{tab:phot}.

Additional 1.2\,mm
photometry of SSG~1 was obtained  using MAMBO in 2005 January for
$\sim$3 hours in photometry mode.  The horn antenna design produces
incomplete sampling of the field, and because there was no jiggling,
SSG~2 was not observed.  The data were reduced using the standard
package ({\sc mopsic}).  

New VLA observations of  the Bo\"otes Deep Field  were
obtained on 2006 April 11.  The VLA was in A configuration, giving a
synthesized beam size of $1\farcs36 \times 1\farcs50$ at 1.5\,GHz.
Wide field imaging mode was used in order to avoid bandwidth
smearing: two 25\,MHz IFs of two polarizations each
with 7 channels per IF. The phase center was located 10\arcsec\ north
of SSG~1 to avoid any possible phase center artifacts. Standard wide
field imaging techniques were employed, including self-calibration
and 3D corrections.  The final image was generated with AIPS
parameter {\sc robust} = 1, giving an rms noise $\approx$10~$\mu$Jy/beam.
Automatic source search and flux density measurements were done with
AIPS task {\sc sad}.  Both SSG~1 and 2 were detected, but neither was
spatially resolved.

Initial {\it Spitzer}/IRAC (\citealt{Fazio2004}) and MIPS
(\citealt{Rieke2004}) data were obtained during surveys of the NOAO
Deep Wide Field in Bo\"otes (\citealt{Eisen2004}).  Significantly
deeper observations were later obtained in 2006 February as part of the
IRAC GTO program (Program ID 520).  The IRAC observations consisted
of six dithered 100\,s frames in each IRAC field of view, covering
the SHARC~II survey area in all four IRAC bands.  Multiplexer bleed
and other detector artifacts were removed by applying the {\it
Spitzer} Science Center's artifact mitigation code to the S14 version
of the automated IRAC pipeline Basic Calibrated Data products.  These
cosmetically enhanced frames were then mosaiced using IRACProc
v4.0beta (\citealt{Schuster2006}) to ensure correct treatment of the
noise for both point and extended sources.  The MIPS observations
consisted of eight 30\,s cycles at 24\,$\mu$m only and covered the
entire 350\,$\mu$m map area.  The data were reduced following
standard procedures (\citealt{Gordon2005}), and source flux densities
were measured with PSF fitting.  The 99\% confidence region of SSG~1
contains two visible-wavelength sources.  The new 24\,$\mu$m imaging
shows that the source chosen as the most likely submm counterpart by
\citet{Khan2005} is responsible for $\sim$75\% of the
24\,$\mu$m emission.

\begin{figure*}
\begin{center}
\epsfig{file=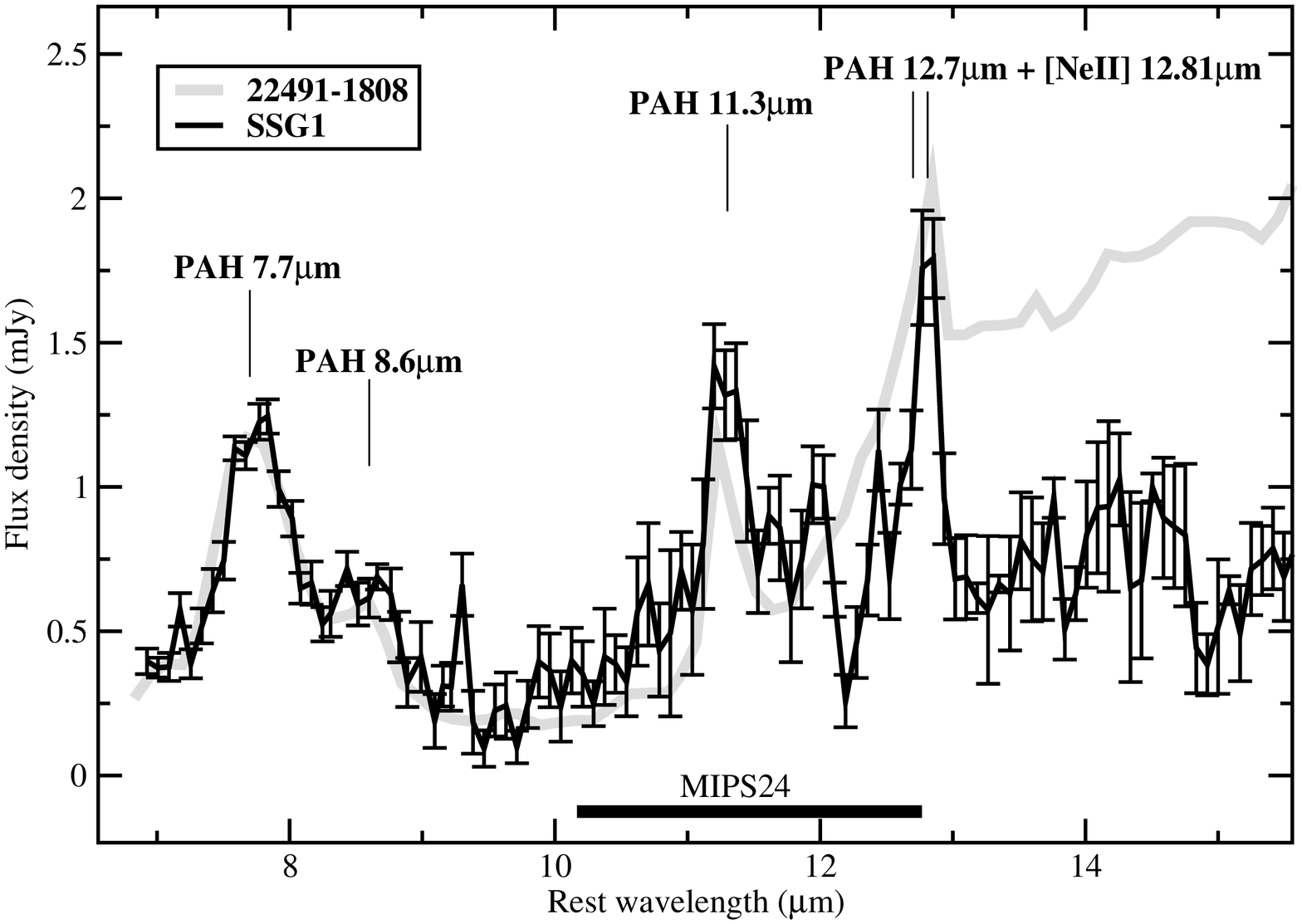,width=4in, angle=270}
\caption{The mid-IR spectrum of SSG~1 as observed by {\it Spitzer} IRS, 
indicated by black line.  One-sigma error bars are shown.
Wavelengths are in the rest frame based on the derived redshift
$z=1.05$. The spectrum of 
IRAS 22491$-$1808, used as a template to derive the redshift, is
shown in grey.  Prominent dust grain
emission features are also indicated. The MIPS 24\,$\mu$m passband is shown by the
heavy black line below the spectrum.
\label{fig:spectrum}}
\end{center}
\end{figure*}

The {\it Spitzer}/IRS observation of SSG~1 was designed according to the
recommended SSC set-up.  For each slit, there were six pointings
along the slit at 24\arcsec\ spacing, all 
centered on the slit in its narrow dimension.  This is equivalent to the normal
point source method except that the target is observed in six
different slit positions instead of the usual two, lessening the
effects of flat-fielding errors and bad pixels.  In 3.1 hours (the
total for the AOR, including peakups and overheads), there were five
cycles of LL1 (19.5--38.0\,$\mu$m) and seven of LL2
(14.0--21.3\,$\mu$m) with 120\,s ramps at each of the six slit
positions.  The data were reduced using the pipeline S15, cleaned
using {\sc irsclean}, and extracted with {\sc smart}
(\citealt{Higdon2004}). Two 
extraction techniques were compared: the first method involved
extracting the spectra from the individual images and taking the
median spectrum; the second method was to align the images, and then
extract the spectrum. There was little difference in the final
spectrum using either technique, and the spectrum is shown in
Figure~\ref{fig:spectrum}.  The most prominent features in the SSG~1
spectrum are due to polycyclic aromatic hydrocarbons (PAH).  For
quantitative analysis, PAH can be measured from integrating the flux
above a baseline, but this method tends to underestimate the flux
density as compared to a profile decomposition method such as the one
used by {\sc pahfit} (\citealt{Smith2006}; see also \citealt{Galliano} for
more detailed consideration of the two approaches).  While both
methods give similar PAH ratios and similar spatial variations of PAH
strengths, absolute flux densities will differ due to assumptions about
the PAH profiles.

\subsection{Counterpart Identification}

\begin{figure*}
\begin{center}
\plotone{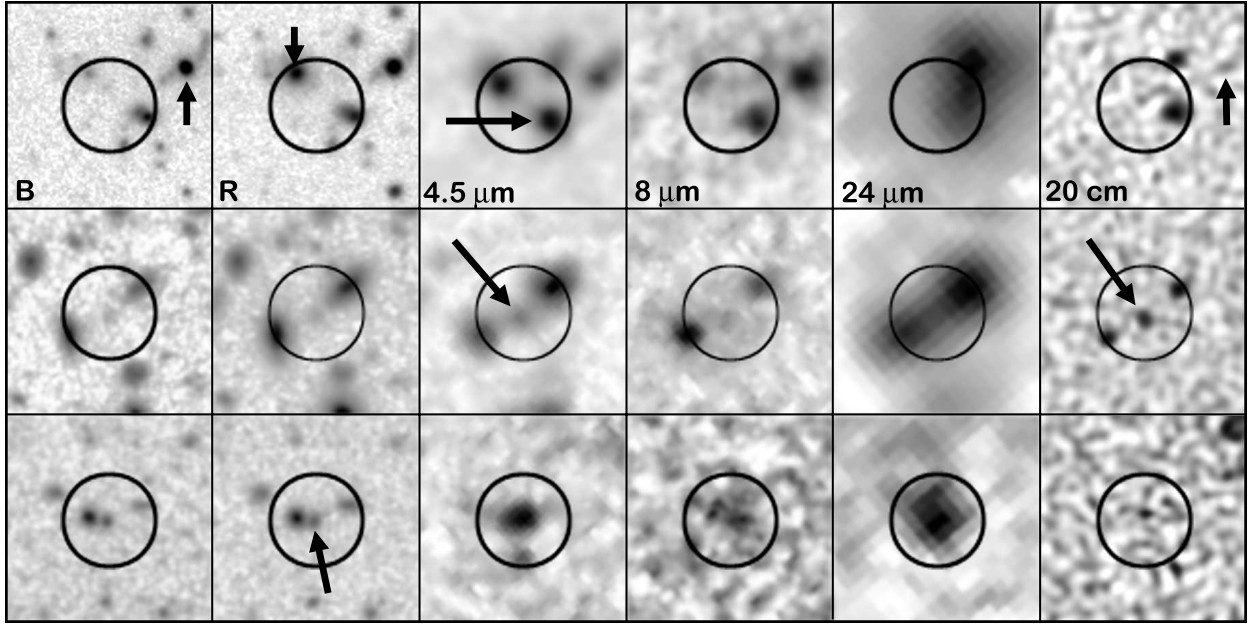}
\caption{Visible, mid-IR, and radio imaging of  LH 350.1 (top), 
LH~350.2 (middle), and LH~350.3 (bottom).  Each image is  20\arcsec\ on
a side with a
9\arcsec\ diameter circle  centered on the SHARC~II 350\,$\mu$m
position.  North is up and east to the left.  Arrows in the LH~350.1
panels indicate the unrelated QSO RDS~054A ($B$ and 20\,cm panels), a
star or low-redshift E galaxy ($R$ panel), and the proposed SMG
counterpart (4.5\,\micron\ panel).  Arrows in the LH~350.2 and 350.3
panels indicate the proposed SMG counterparts.
\label{fig:lhstamp}}
\end{center}
\end{figure*}

The additional deep 24\,$\mu$m and 1.4\,GHz imaging of the Bo\"otes
field yielded counterparts in these bands for SSG~1 and 2 with low
probabilities of chance association (\citealt{Khan2007}). The more
precise positions were then 
used to identify optical and mid-IR counterparts.  Finding
counterparts for the three sources in the Lockman Hole is not
straightforward: Figure~\ref{fig:lhstamp} shows thumbnail images
of the fields at various wavelengths.  As can be seen, there exist
multiple radio detections for two of the 350\,$\mu$m objects.
However the {\it most likely} counterparts can be determined with the
following rationale:

\begin{itemize}

\item LH 350.1: the 1.4\,GHz image shows two potential counterparts,
but only the southern one is within the 9\arcsec\ SHARC~II beam.  In
the visible images this appears to be a blend of two sources.  The
photometric redshift for this southerly source was obtained using
{\sc hyper-z} (\citealt{Bolzonella2000}) in conjunction with the
$BRIzJK$[3.6][4.5] photometry, similar to the approach of
\citet{Dye2008}. The best-fitting solution is a starburst template at
$z=1.19^{+0.33}_{-0.14}$.  The northern 1.4\,GHz object is not associated with
any visible-light detection, although it is prominent in the IRAC and
MIPS images.  Those images also show two more sources: an easterly
object whose SED suggests it is a star or a low redshift elliptical
galaxy, and well to the west, the QSO RDS 054A at $z=2.416$
(\citealt{Schmidt1998}), which is not expected to make a strong
contribution to the 350\,$\mu$m emission.

\item LH 350.2: the 1.4\,GHz image 
shows three sources within the beam.
However, the two outermost objects are well-resolved in the visible
imaging, suggesting that they are likely to be relatively
nearby.  Photo-z fitting puts both at $z\sim 0.4$, and SED template
fitting puts a contribution to the 350\,$\mu$m emission of $<$20\%
(assuming a dust temperature, $T_d >30$\,K).  Therefore, the very red
middle object is likely to be the predominant source of the
350\,$\mu$m emission with a best-fitting photo-z of
1.21$^{+0.13}_{-0.20}$.

\item LH 350.3: there is a strong 24\,$\mu$m and weak 1.4\,GHz source
within 6\arcsec\ of the 350\,$\mu$m position.  The $B$ through
4.5\,$\mu$m images suggest that the mid-IR flux is a combination of
two sources with the radio position lining up on the redder, western
object.  Best-fitting photometric redshifts are
$z=2.47^{+0.51}_{-0.97}$ and $z=0.56^{+0.28}_{-0.39}$ for the western
and eastern objects respectively.  There remains the 
possibility of a merging system because the secondary solution for the
eastern source is $z=2.60^{+0.32}_{-0.33}$.  Based on the radio
identification, the western object is taken as the main source of the
submillimeter emission.

\end{itemize}

\label{subsec:lh}



\label{subsec:sed}

\section{Constraints on the photometric redshift and thermal emission}

\begin{figure*}
\begin{center}
\epsfig{file=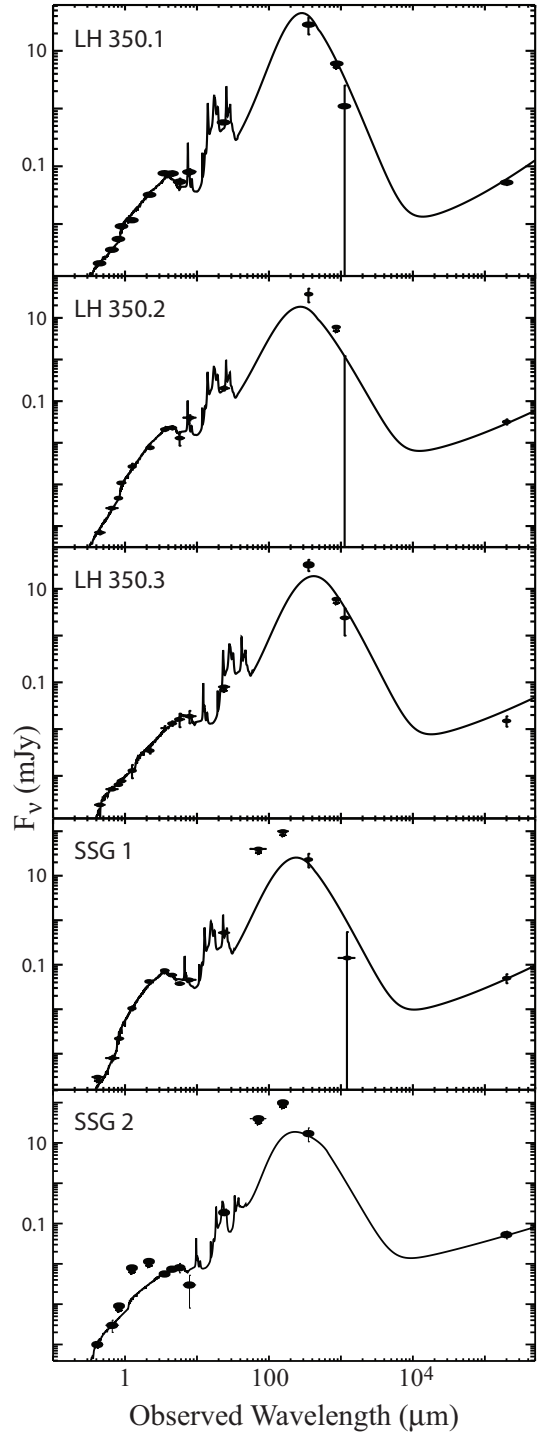, height=7.5in}
\caption{Best-fitting  SED templates from STARDUST2.  Wavelengths are
in the observed frame.
\label{fig:sedtemplates}}
\end{center}
\end{figure*}

The multiwavelength photometry in Table~\ref{tab:phot} was used to
derive photometric redshifts and thermal emission parameters.
Following the same approach as \citet{Khan2005}, the {\sc stardust2}
template fit was used (\citealt{Chanial2009}) to simultaneously
obtain the photometric redshift $z_{\rm phot}$, dust temperature $T_d$,
and the 8--1000\,$\mu$m IR luminosity $L_{\rm IR}$.  Templates were based
on local starbursts, and $\chi^2$ was minimised through a
Levenberg-Marquardt technique.  For SSG~1, the redshift obtained for
the IRS spectrum in Figure~\ref{fig:spectrum} was used in the fit,
but $z_{\rm spec}= 1.05$ agrees well with the $z_{\rm phot}=
1.03$ from the template fitting.
The best-fit parameters are given in Table \ref{tab:stardust2}, and
the best-fitting templates shown in Figure~\ref{fig:sedtemplates}.

For SSG~1, SSG~2, and LH 350.3 the 350\,$\mu$m flux density is well
fit by the template, but in the cases of LH 350.2 and 3, the quality
of the fit at 350\,$\mu$m is clearly affected by the longer
submillimeter data, which constrain the slope on the
Rayleigh-Jeans side of the spectrum.  Additionally, the 24\,$\mu$m
flux density is useful in determining which side of the peak the
350\,$\mu$m datapoint lies.  Usually the mid-IR is not used to
constrain the restframe far-IR continuum emission because specific dust
grains and AGN can have a major effect on mid-IR flux densities.
Usage of the MIPS band was checked for a sample of SMGs (from
\citealt{Coppin2008}) by comparing template fits using measurements
at 350\,$\mu$m, 
850\,$\mu$m, and radio with measurements at 24\,$\mu$m, 350\,$\mu$m,
and radio plus long-submillimeter limits. The two approaches agree
within the {\sc stardust2} uncertainties.  This indicates our thermal
parameters are not noticeably biased by the inclusion of the mid-IR
data. (See \citealt{Chanial2009} for more details.)

For LH 350.3, the lack of a defining 1.6\,$\mu$m rest-frame stellar
continuum feature makes a significant AGN contribution plausible,
although the relatively low radio flux density would argue against
this.  Because of the absence of AGN templates in {\sc stardust2},
the LH~350.3 data were also fit using another routine with a mixture
of AGN and starburst templates (\citealt{Negrello2009}).  In this
template fit, the favoured solution is a starburst at $z_{\rm
phot}=3.8$ (but without including the radio flux density).  Therefore
we consider LH 350.3 to be, like the rest of this sample,
predominantly star-forming.  Given the photometric redshift estimates
from hyper-z, {\sc stardust2}, and \citet{Negrello2009}, it is
possible that LH 350.3 is the most distant  galaxy
so far discovered at  short-submillimeter wavelengths.

\subsection{Energy source diagnostics}

As can be seen in Figure~\ref{fig:sedtemplates}, all five sources are
well fit in the mid-IR by templates that include features associated
with star formation, \ie, a rest-frame 1.6\,$\mu$m stellar continuum
bump and apparent PAH emission.  For SSG~1, the presence of PAH is
confirmed by the mid-IR spectrum, shown in Figure~\ref{fig:spectrum}.
An emission line from [\ion{Ne}{2}]
is likely present though blended with possible PAH emission at
12.7~\micron.  The dust continuum is relatively flat until (rest wavelength)
$\sim$15.5\,$\mu$m.  Although the 9.7\,$\mu$m silicate absorption is
relatively weak, it is within the range seen in local IR-luminous
galaxies.  Table \ref{tab:spectrum} gives the emission feature fluxes, and a
standard template (IRAS 22491$-$1808, an IR-luminous galaxy with
moderately strong PAH emission) was used to obtain an overall
best-fitting redshift from the spectral features of 1.05$\pm$0.01.

The values of $L_{\rm PAH}$ for the 7.7 and 11.3\,$\mu$m rest-frame
emission (6.8 and 2.0$\times$10$^{9}$\,\Lsun) are consistent with the
$L_{\rm PAH}$ vs $L_{\rm IR}$ distribution of \citet{Pope2008} for a
sample of 850\,$\mu$m-selected SMGs. (This comparison used the
integrated line flux densities with the $L_{\rm IR}$ given in
Table~\ref{tab:stardust2}). The 7.7\,$\mu$m PAH luminosity also
provides an estimate of the star formation rate (SFR) of
340\,\Msun\,yr$^{-1}$ following the relation given by \citet{Weedman2008}.
Given the strong star-forming features in the mid-IR spectrum, this
compares favorably with the SFR derived from $L_{\rm IR}$
of 136\,\Msun\,yr$^{-1}$ following the relation given by
\citet{Kennicutt1998}.


%
\section{Discussion}

The different survey depths make it complicated to produce fair
statistics for number counts.   The  Lockman Hole surveys 
\citep{Laurent2006,Coppin2008} were targeted at known sources, and
sensitivity for serendipitous sources varied with distance from the
target source.  Even the blank field survey in Bo\"otes had
sensitivity varying with location. (See Fig.~3
of \citealt{Khan2007}.)  However, simply considering the flux
densities of the five detections and the total survey area
(15.1\,arcmin$^2$)\footnote{5.8\,arcmin$^2$ for the Bo\"otes
observations; for LH 350.1 and 2 \citep{Laurent2006}
the map area is assumed to be that
of the arguably similar SHADES observation of LH 350.3 of
3.1\,arcmin$^2$; \citealt{Coppin2008})}, we estimate resolving
$\sim$20\% of the 350\,$\mu$m contribution to the cosmic infrared
background (CIB; \citealt{Fixsen1998}) at $S>17$\,mJy. This fraction
is reasonable given that the bulk of the background will be generated
by galaxies near the break in the differential counts.  It is also 
consistent with the 30\% estimate at $S>13$\,mJy from the source
counts of \citet{Khan2007}.

A lower limit on the star formation rate density (SFRD)
can be found using the thermal parameters in Table~\ref{tab:stardust2}
and the comoving volume in $1<z<3$.  The  conversion from far-IR
luminosity to SFR is given by \citet{Kennicutt1998}.
The result  for this epoch is
0.017\,M$_\odot$\,yr$^{-1}$\,Mpc$^{-3}$, much lower
than the values presented by \citet{Hopkins2006}.  This implies the
bulk of the 350\,$\mu$m contribution to the star formation rate
density is from less luminous infrared galaxies and normal galaxies,
consistent with observations of IR-luminous galaxies in different
bands (e.g., \citealt{LeFloch2005}) and the predictions from source
count models (e.g., \citealt{PK2009}).

\begin{figure*}
\begin{center}
\epsfig{file=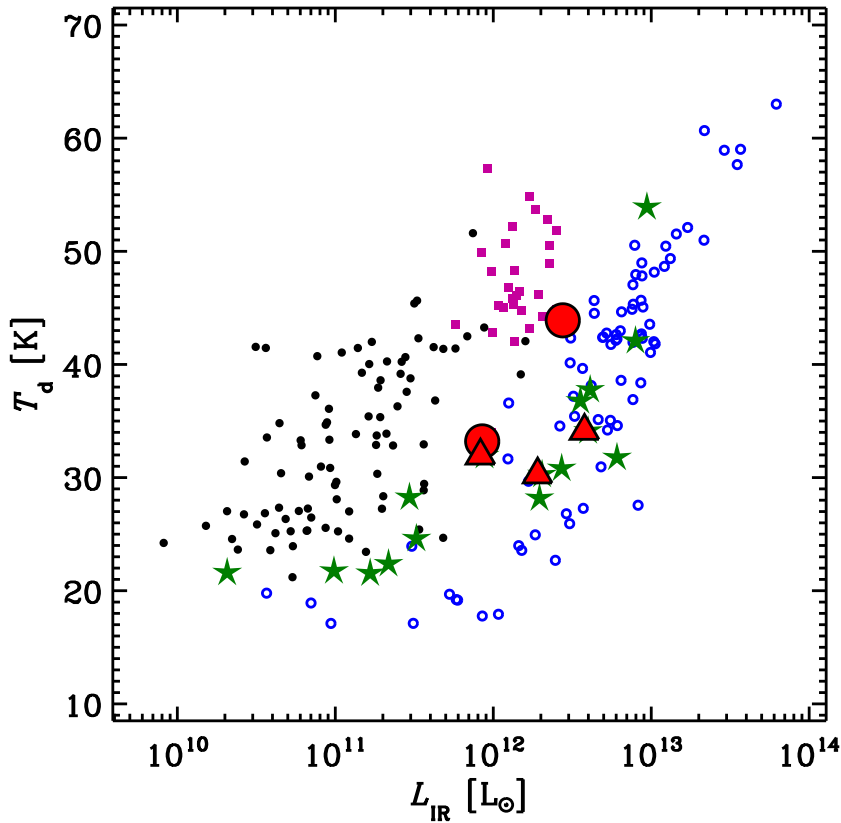,width=5in}
\caption{The IR luminosity-dust temperature distribution for various
IR-luminous galaxy samples.  For the 350\,$\mu$m-selected galaxies,
the red circles  show SSG~1 and SSG~2,  and triangles show LH 350.1,
350.2, and 350.3.  Green stars represent the 850 and
1100\,$\mu$m-selected SMGs for which there is 350\,$\mu$m follow-up
photometry.   This is compared with 850\,$\mu$m-selected SMGs from
\citet{Chapman2005} (open blue circles) and {\it IRAS} galaxies from
\citet{Dunne2000} (filled black squares) and \citet{Clements2009}
(purple squares).
\label{fig:lirtd}} 
\end{center}
\end{figure*}

Our 350\,$\mu$m-selected galaxies can be compared with sources
selected at longer submillimeter wavelengths.  Suitable samples that
include 350 and 24\,$\mu$m photometry come from \citet{Coppin2008},
\citet{Laurent2006}, and \citet{Kovacs2006}.  The photometric
redshifts and thermal parameters for these SMGs were again found
using {\sc stardust2}.  Where there is no spectroscopic redshift, the
photo-z was obtained from optical and mid-IR magnitudes given by
\citet{Dye2008} and \citet{Clements2008}.  Following the $L_{\rm
IR}$--$T_d$ distribution by \citet{Blain2004}, there appear to be two
trends separating the predominantly local {\it IRAS} galaxies from
the higher redshift SMGs. (See Fig.~\ref{fig:lirtd}.) Compared with
those selected at
850 and 1100\,$\mu$m, the 350\,$\mu$m-selected sources have similar
dust temperatures and give no indication of temperature bias
according to selection wavelength. Deeper 800--1000~\micron\ imaging
(e.g., with ALMA --- \citealt{Wootten2009}) of the
350~\micron-selected population would  
provide a more direct direct comparison of any selection biases.

\begin{figure*}
\begin{center}
\epsfig{file=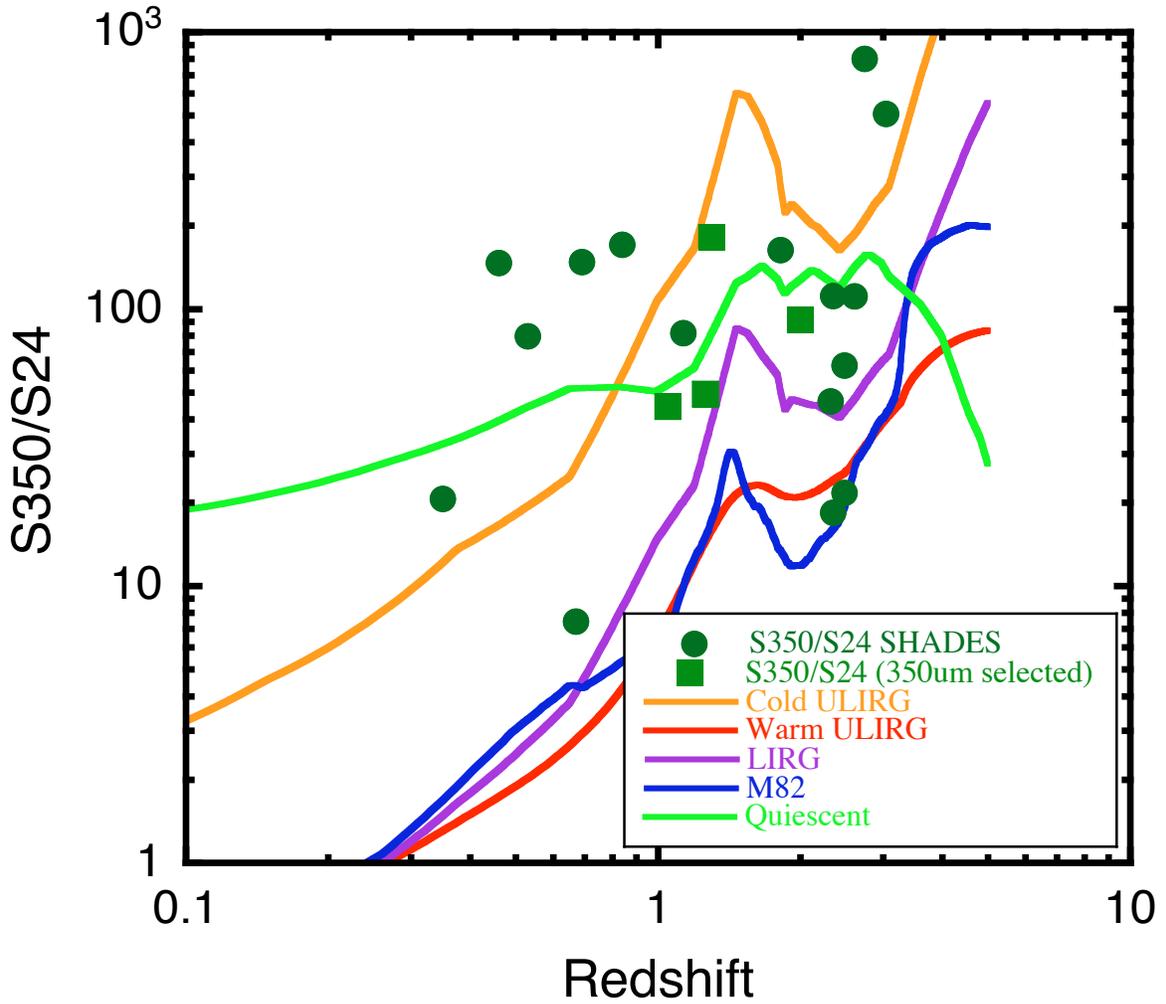,width=6in}
\caption{The 350\,$\mu$m:24\,$\mu$m flux density ratio
as a function of redshift for various SMG
samples as indicated in the legend. Lines show the same ratio for
local SED templates from \citet{Efstathiou2000}.
\label{fig:col}} 
\end{center}
\end{figure*}

Mid-IR surveys to probe the $1<z<3$ galaxy population
have the advantage of better angular resolution,
aiding counterpart identification (and hence more reliable redshift
estimation), but a significant disadvantage is that without further
far-IR observations, constraints on the
rest-frame far-IR thermal emission must be indirectly inferred.  The
350:24 flux density--redshift distribution for the SHARC~II
350\,$\mu$m-selected SMGs compared  with the 850 and 1100\,$\mu$m-selected
SMGs is shown in Figure~\ref{fig:col} alongside colors from various
templates from \citet{Efstathiou2000}.  The overall distribution
suggests that all samples are drawn from the same IR-luminous galaxy
population, and the good agreement with the templates
suggests these are predominantly luminous and
ultraluminous infrared galaxies.\footnote{$10^{11}<L_{\rm IR} <
10^{12}$\,\Lsun\ and $L_{\rm IR}>10^{12}$\,\Lsun\ respectively.}

\begin{figure*}
\begin{center}
\epsfig{file=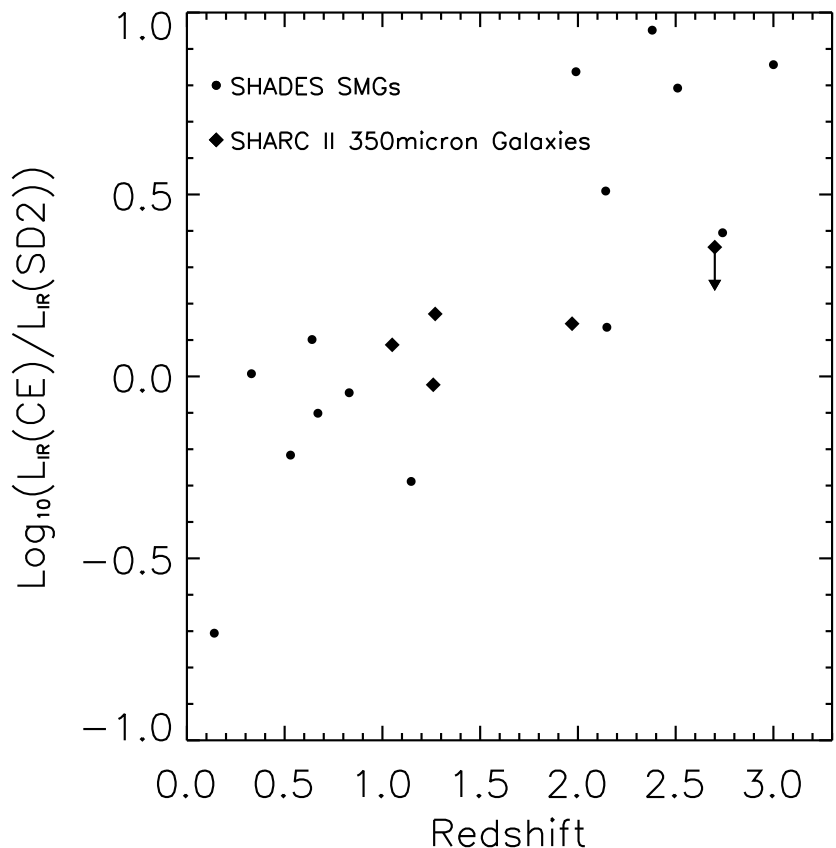,width=5in}
\caption{The ratio of best-fitting IR (8--1000\,$\mu$m) luminosities
obtained from the CE and STARDUST2 fitting, as a function of
redshift.  The diamonds 
denote the 350\,$\mu$m-selected galaxies, and filled circles denote
the 850 and 1100\,$\mu$m-selected galaxies. The diamond with arrow
denotes an upper limit.   Sources with redshifts beyond the useable
range for the CE model ($\sim$3) are excluded from consideration. 
\label{fig:ir24}} 
\end{center}
\end{figure*}

Studies of infrared galaxies are often limited by unavailability of
data at wavelengths longer than 24\,\micron.  In such cases, the IR
luminosity can be estimated using only the 24\,$\mu$m flux density
and various SED templates (\citealt{CharyElbaz}) combined with an
estimated redshift. For the SMG
samples, the thermal emission estimates found this way compare
surprisingly well with those from {\sc stardust2} for $z\la 1$ as
shown in Figure~\ref{fig:ir24}.  For these galaxies, the dominant
errors on both luminosity predictors will arise from the photo-$z$
fitting; these errors are highly non-Gaussian and non-trivial to determine.
To some extent, the agreement in Figure~\ref{fig:ir24} can
be attributed to small number statistics and selection.
The overall variation in the ratio over a wider redshift range could
also imply a heterogenous SMG population.  This would be consistent
with the color in Figure~\ref{fig:col} but with a varying degree of
AGN contribution.  Another condition on this sample is the
requirement of bright optical, mid-IR, and most importantly radio
identifications; at high redshift, an AGN will boost the radio flux density
and the emission in other bands (notably in the mid-IR), allowing
this type of analysis to be performed.  However none of the sources
in the sample appears to be strongly affected by an AGN.  At higher
redshifts multiwavelength analysis becomes much more difficult.  
Some of the higher-$z$ 850/1100\,$\mu$m galaxies
could simply have incorrectly identified 24\,$\mu$m counterparts
(see, e.g., \citealt{Younger2007}).

The restframe 12\,$\mu$m flux density has been suggested to be a
``pivot point'' for SEDs (\citealt{Spinoglio1995}; see also
\citealt{Elbaz2002} ), giving a fixed $L_{12}/L_{\rm FIR}$ for both
starbursts and AGN.  This correlation was determined from local {\it
IRAS} sources, but at $z\sim1$, the rest frame 12\,$\mu$m is shifted
into the 24\,$\mu$m band, and thus the relation at this redshift can
be examined directly by our sample. We expect this relation to hold
if the populations are similar. An observed difference in the populations is
that  distant mid-IR sources
show stronger dust grain emission features around 12\,$\mu$m
compared to the local IR-luminous galaxy population.   
This difference could be due to differing AGN fractions
\citep{Wu2009}.

\begin{figure*}
\begin{center}
\epsfig{file=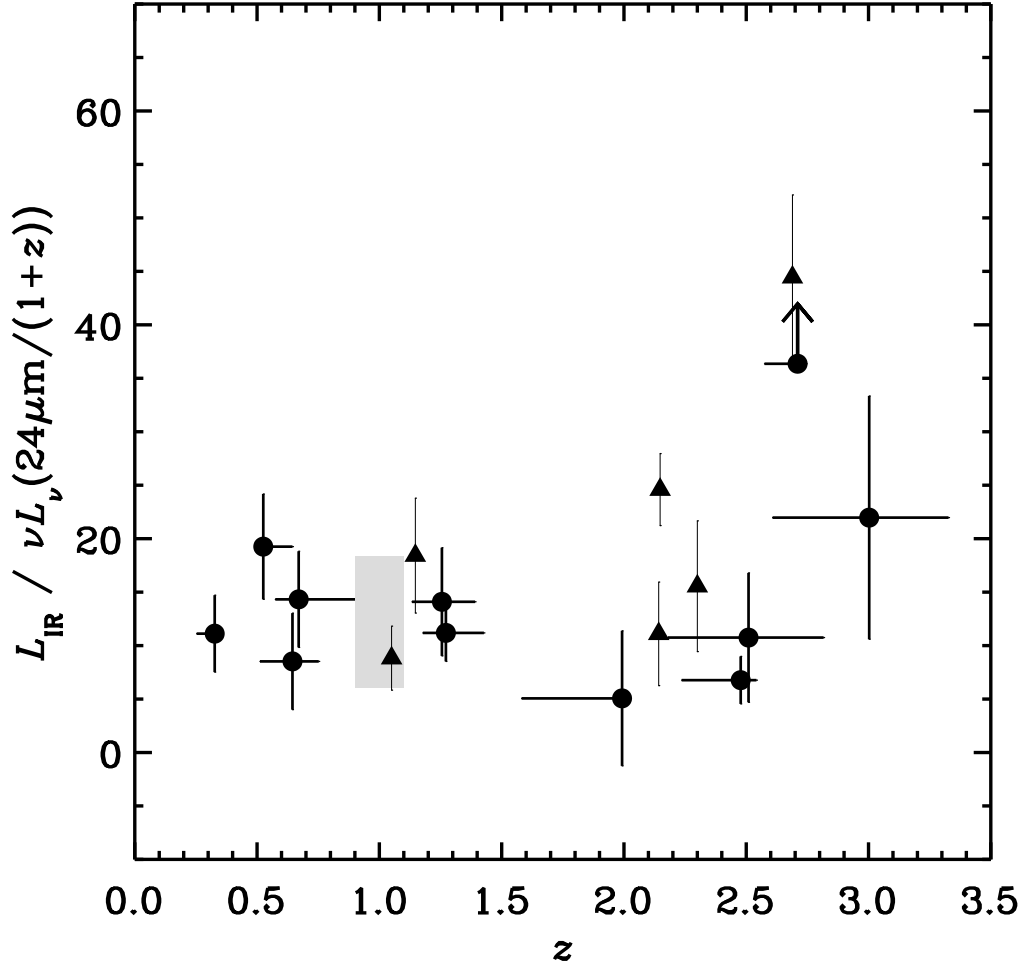,width=6in}
\caption{Ratio of  far-IR luminosity to monochromatic luminosity at
observed wavelength
24\,\micron\  as a function of redshift.  Data are shown for the
350, 850, and 1100\,$\mu$m-selected SMGs.  Numerator of the ratio is
the rest frame far-IR (8--1000\,$\mu$m) luminosity derived from the
STARDUST2 template fitting.
The denominator is the rest frame luminosity for the  24\,$\mu$m
observation, i.e., monochromatic luminosity $\nu L_\nu$ at rest frame
wavelength $24\,\micron/(1+z)$.   At $z\approx1$,
observed 24\,\micron\ corresponds to rest 12\,\micron, which as
\citet{Spinoglio1995} suggested gives a good measure of total
luminosity.  The shaded region on the plot shows the
\citeauthor{Spinoglio1995} data for local luminous infrared galaxies
as they would appear at 
$z=1$.  Filled triangles denote sources with spectroscopic redshift,
while filled circles denote sources with photometric redshifts from
the STARDUST2 template fit.
\label{fig:spin}} 
\end{center}
\end{figure*}

Figure~\ref{fig:spin} shows the ratio of 
far-IR luminosity  to observed {\it Spitzer}/MIPS
24\,$\mu$m monochromatic luminosity  as a
function of redshift.  The ratio 
flattens out at $z\sim1$ as the rest frame 12\,$\mu$m emission is
shifted into the MIPS 24\,$\mu$m band, but the dispersion increases
at higher redshift.  This suggests that the local empirical
far-IR/12\,$\mu$m relation could apply to earlier epochs where the
galaxy is undergoing rapid evolution, making rest-frame 12\,$\mu$m
flux densities a better tracer of IR luminosity \citep{Spinoglio1995}
than other mid-IR
emission features, e.g., 7.7\,$\mu$m PAH selected at $z\sim2$.  
At $z\sim1.4$ the silicate absorption feature is redshifted into
the MIPS 24\,$\mu$m band and  affects the flux density
measurement. This effect is emphasized in Figure ~\ref{fig:col}, where
an excess in the 350/24\,$\mu$m colours is evident in the model
SEDs, and
galaxies in the redshift range where the silicate feature would
cause a deficit in the 24\,$\mu$m flux ($1.3<z<1.6$) are notably absent from
Figure \ref{fig:spin}.

Various studies have suggested that the far-IR luminosity can be
deduced solely from mid-IR observations of
galaxies. \citet{Bavouzet2008} have shown that the far-IR luminosity
of dusty galaxies observed with {\it Spitzer} correlate closely with the
corresponding 8 and 24\,$\mu$m mid-IR luminosities. These studies
were limited to redshifts $z\la1$ by
the necessity to use  {\it Spitzer} bands at 70
and 160\,$\mu$m  to calculate the far-IR luminosity.  The
mid-IR to far-IR luminosity relation was tentatively
extended to higher redshifts ($z\sim2$) by stacking the far-infrared
fluxes in the long wavelength {\it Spitzer} bands, and these stacked
results were used by \citet{Caputi2007} to calculate the bolometric
luminosity function at $z\approx2$. In contrast, our submillimeter
observations directly access the high redshift
Universe without the need for stacking images to obtain far-infrared
fluxes.  From Figure~\ref{fig:spin} we observe a tight correlation
between the far and mid-IR flux densities in our sample between
$1<z<2$, but there is a large dispersion in the infrared colours
at $z>2$. For $z>2$, the 24\,$\mu$m band is sampling  rest
wavelengths shorter than 8\,$\mu$m, and the far-IR
luminosity cannot be accurately predicted from such observations.  In
order to confirm the 
mid-IR/far-IR relation for $1<z<2$ and to test the
relation to higher redshifts, large samples of SMGs will be
required. Such samples will be available following the large surveys
planned with Herschel SPIRE (at 250, 350, and 500\,$\mu$m;
\citealt{Griffin2008}) .

\section{Conclusion}

The discovery of the first 350\,$\mu$m-selected galaxies using
SHARC~II offers a different insight into IR-luminous galaxies than
those selected in either the mid-IR or in longer submillimeter bands.  The
present sample comprises IR-luminous galaxies at $1<z<3$ and resolves
$\sim$20\% of the 350\,$\mu$m background at $S>17$\,mJy.  The implied
lower limit on the star formation rate density suggests that the bulk of
the 350\,$\mu$m contribution is from galaxies of lower IR luminosity
than the ones detected at present flux densities.

The combination of 350\,$\mu$m and mid-IR flux densities allows us to
examine the relation of the mid and far-IR emission in dusty galaxies
at high-redshift.  There is surprisingly good agreement at $z<2$,
with the flattening of the IR:$24\,\micron/(1+z)$ luminosity ratio at
$z\sim1$ evidence 
for the local empirical relation of rest frame 12\,$\mu$m IR
luminosity tracing IR luminosity applying to distant IR-luminous
galaxies.  The wider dispersion at $z>2$ suggests a limit to which
the mid-IR can be used as a proxy for far-IR emission.  We also find,
by comparing our 350\,\micron-selected sample to samples chosen at
850\,\micron\ and millimeter wavelengths, no evidence for a cold
color temperature bias in SMGs.  These results will be further
tested through surveys with {\it Herschel}-SPIRE, which will detect
SMGs in unprecedented numbers.

The present 350\,\micron-selected sample, with $15<S_{350}<40$\,mJy,
is near the SPIRE 350\,$\mu$m blank-field detection limit (e.g.,
\citealt{PK2009,Franceschini2009}) and also the detection limits of
current longer-wavelength
submillimeter instruments. Counterpart identification is
difficult even with the available angular resolution and
will be far more difficult for space-based surveys
with smaller telescopes. The achievable sensitivity and
resolution of ground-based detectors will therefore provide a necessary
complement to the space-based surveys, especially in characterising
the nature of the faint SMG population.  Whether the first
350\,$\mu$m-selected galaxies are typical of the
population will be probed through the combination of surveys with
SPIRE and future surveys from the ground (e.g., SCUBA 2--450\,$\mu$m;
\citealt{Holland2006}).



%

\acknowledgements

We thank Tom Phillips and the CSO for observing time and assistance
during our runs and Darren Dowell, Colin Borys, Attila Kovacs, Rick
Arendt, Dave Chuss, and Bob Silverberg for providing observing and
data reduction support.  We thank Glenn Laurent and Kristen Coppin
for advice on their 350\,$\mu$m Lockman Hole follow-up surveys, Mark
Brodwin and Yen-Ting Lin for assistance in compiling the Bo\"otes
optical and near-IR photometry, and Rob Ivison for providing the radio
maps for the Lockman Hole sources.

The Caltech Submillimeter Observatory is supported by NSF contract
AST-0229008.
This work is based in part on observations made with the Spitzer
Space Telescope, which is operated by the Jet Propulsion Laboratory,
California Institute of Technology under a contract with NASA.
This work is based in part on data collected at Subaru Telescope, which is
operated by the National Astronomical Observatory of Japan. 
The Subaru data are those used by \citet{Dye2008}.
This research has made use of the NASA/IPAC Extragalactic Database
(NED), which is operated by the Jet Propulsion Laboratory, California
Institute of Technology, under contract with the National Aeronautics
and Space Administration.
This research has made use of the SIMBAD database, operated at CDS,
Strasbourg, France.   
This research draws upon data provided by Buell Jannuzi \& Arjun Dey
and by Richard Elston \& Anthony Gonzalez (\citealt{Elston2006}) as
distributed by the NOAO Science Archive. NOAO is operated by the
Association of Universities for Research in Astronomy (AURA),
Inc.\ under a cooperative agreement with the National Science
Foundation.
The UKIDSS project is defined by \citet{Lawrence2007}. UKIDSS uses
the UKIRT Wide Field Camera (WFCAM; \citealt{Casali2007}).  We 
used data from the fourth data release (DR4).

S.A.K.\ is partially supported through FONDECYT Proyecto 1070992 and
ALMA-Conicyt.  Support for E.L.F.'s work was provided by NASA through
the Spitzer Space Telescope Fellowship Program.

Facilities: \facility{CSO, Spitzer, UKIRT, Subaru, VLA}



\clearpage 

\clearpage 

\begin{table}
\begin{center}
\caption{\protect\centering The first five 350\,$\mu$m-selected
galaxies}
\begin{tabular}{cll} \tableline\tableline
Name [IAU] (nickname) & \multicolumn{2}{c}{Position [VLA] (J2000)}\\ 
Other names & \multicolumn{2}{c}{Position [350\,$\mu$m] (J2000)} \\
\tableline
SMM J143206.65+341613.4 (SSG 1) & 14 32 06.58 & +34 16 11.9\\
  & 14 32 06.65 & +34 16 13.4\\ 
SMM J143206.11+341648.4 (SSG 2) & 14 32 06.04 & +34 16 46.7\\
& 14 32 06.11 & +34 16 48.4\\ 
SMM J105308.3+571501 (LH 350.1) & 10 53 07.89 & +57 15 00.3 \\
SHARC~II Source 3 &  10 53 08.3 & +57 15 01\\
SMM J105232.3+572448 (LH 350.2) & 10 53 32.26 & +57 24 47.4 \\
SHARC~II Source 4 & 10 52 32.3 & +57 24 48\\
SMM J105243.2+572309 (LH 350.3) & 10 52 43.17 & +57 23 09.7 \\
LOCK 350.1 & 10 52 43.2 & +57 23 09\\ 
\tableline
	\end{tabular}		
\\\smallskip
{VLA positions are for identified VLA counterparts.
The 350\,$\mu$m position  is used for the IAU name.} 
\label{tab:lh}
\end{center}
\end{table}

\clearpage
\begin{table}
   \begin{center}
\caption{\protect\centering
Photometry for counterparts of 350\,\micron-selected galaxies}
\begin{tabular}{cccccc} \tableline\tableline
$\lambda_{obs}$  & SSG 1 & SSG 2 & LH 350.1 & LH 350.2 & LH 350.3 \\  
 			& 	& 	& 	& 	& \\ \tableline
$B$ [$\mu$Jy]                & $<$0.3   & 0.1$\pm$0.02 & 2.1$\pm$0.020 & 0.07$\pm$0.007 & 0.24$\pm$0.009    \\  
$R$                       & 0.8$\pm$0.1 & 0.3$\pm$0.1  & 3.6$\pm$0.033 & 0.27$\pm$0.013 & 0.52$\pm$0.014 \\ 
$I$             		  & 2.2  $\pm$ 0.1 & $<$ 0.9   & 5.5 $\pm$ 0.051 & 0.47 $\pm$ 0.017  & 0.65 $\pm$ 0.018\\
$z$         & \nodata & \nodata & 9.2 $\pm$ 0.085 & 1.09 $\pm$ 0.040 & 0.78 $\pm$ 0.036\\
$J$       	            & 10.5 $\pm$ 1.1 & $<$ 7.9 &  11.7 $\pm$ 0.65 & 2.7 $\pm$ 0.45 & 1.3 $\pm$ 0.41\\  
$K$        	   & 42.0 $\pm$ 1.4  & $<$ 11.5   & 32.2 $\pm$ 0.89 & 7.6 $\pm$ 0.56  &  $<$1.5  \\
3.6\,$\mu$m         &  73.8 $\pm$ 0.3 & 5.6 $\pm$ 0.3  &       75.5 $\pm$ 7.0 & 21.1 $\pm$ 2.0 &  10.5 $\pm$ 1.1 \\
 	4.5\,$\mu$m         &  58.5$\pm$0.5 &  7.4$\pm$0.5 &  74.8 $\pm$ 6.9 & 23.0 $\pm$ 2.2 & 13.2 $\pm$ 1.5  \\
5.8\,$\mu$m         &  37.7$\pm$2.1  & 8.1$\pm$2.1 & 54.2 $\pm$ 7.0  & 13.0 $\pm$ 4.5 & 16.2 $\pm$ 5.2 \\
8.0\,$\mu$m         &  45.8$\pm$2.2  & 3.0$\pm$2.2 &   80.7 $\pm$ 8.6 & 40.1 $\pm$ 6.0 & 18.9 $\pm$ 5.6 \\
24\,$\mu$m          &  523.8$\pm$58.7   & 187.2$\pm$29.4  &  576 $\pm$ 13\tablenotemark{a}  & 204 $\pm$ 24\tablenotemark{a}  &  $<$80.2\tablenotemark{b}\\
70\,$\mu$m [mJy]         & $<$ 40       &  $<$ 40 & \nodata & \nodata & \nodata\\
160\,$\mu$m         & $<$ 100 & $<$ 100 & \nodata & \nodata & \nodata\\ 
350\,$\mu$m	    & 23.2$\pm$7.9 & 17.1$\pm$6.4  & 28.4$\pm$9.2 & 37.0$\pm$13.4 & 32.8$\pm$8.9  \\
850\,$\mu$m  	 & \nodata & \nodata & $<$6 &  $<$6 & $<$6 \\
1100\,$\mu$m  & \nodata & \nodata & 1.1$\pm$1.4 & -0.2 $\pm$1.4 & 2.4$\pm$1.4   \\
1200\,$\mu$m	    &  0.142$\pm$0.4  & \nodata & \nodata & \nodata& \nodata\\
20\,cm [$\mu$Jy]             & 49.8$\pm$11.4  &  53.1$\pm$11.4 & 52.5$\pm$5.2   & 32.0$\pm$4.6 & 15.8$\pm$4.8 \\ 
\tableline
\end{tabular}		
\tablenotetext{a}{Flux density is deblended using [3.6]\,$\mu$m positions.}
\tablenotetext{b}{Flux density cannot be deblended, used as a limit instead.}
\\ \smallskip
{For non-detections, the flux density at the radio position is used.
Upper limits are  3$\sigma$.  These data are used in the
STARDUST2 SED template fitting.}
\label{tab:phot}
\end{center}
\end{table}

\clearpage
\begin{table}
\begin{center}
\caption{\protect\centering
Best-fitting parameters from STARDUST2 template fitting.}
\begin{tabular}{ccccc}\tableline\tableline
Nickname &  $\log (L_{IR}(SD2)/L_{\odot})$ [$L_{IR}(CE)$]\tablenotemark{a} & $T_{dust}(SD2)$ & $z_{phot}(SD2)$ &  $S_{850}(SD2)$\tablenotemark{b} [mJy]\\ \tableline
SSG~1     & 11.9$^{+0.1}_{-0.1}$ [12.0] & 33.2$^{+2.3}_{-2.3}$ & 1.03$^{+0.16}_{-0.14}$\tablenotemark{c}  & 2.4$\pm$0.8\\
SSG~2     & 12.4$^{+0.1}_{-0.1}$ [12.6] & 43.9$^{+2.4}_{-9.2}$ & 1.97$^{+0.8}_{-0.3}$ & 2.7$\pm$0.7\\
LH 350.1 & 12.3$^{+0.10}_{-0.13}$ [12.5] &  30.3$^{+1.5}_{-1.2}$ &  1.27$^{+0.16}_{-0.09}$ & 6.7$\pm$2.7 \\
LH 350.2  & 11.9$^{+0.13}_{-0.24}$ [11.9] &  31.9$^{+3.4}_{-2.8}$ & 1.26$^{+0.14}_{-0.12}$ & 2.4$\pm$0.6\\
LH 350.3  & 12.6$^{+0.13}_{-0.24}$ [$<$13.0] & 34.1$^{+2.4}_{-1.8}$ & 2.71$^{+0.03}_{-0.14}$  & 7.6$\pm$2.0\\ 
\tableline
\end{tabular}
\label{tab:stardust2}
\end{center}
\tablenotetext{a}{The luminosity from the Chary-Elbaz template fit,
$L_{IR}(CE)$, is  obtained from the 24\,$\mu$m flux density and
the STARDUST2 photometric redshift 
$z_{phot}(SD2)$ (or spectroscopic redshift for SSG~1).}
\tablenotetext{b}{Predicted 850\,$\mu$m flux densities from the
STARDUST2 fit.}
\tablenotetext{c}{Spectroscopic redshift for SSG~1 is $1.05 \pm 0.01$.}
\smallskip
{ } 
\end{table}

\clearpage
\begin{table}
\begin{center}
\caption{\protect\centering Features in the SSG~1 spectrum}
\begin{tabular}{cccc}\tableline\tableline
Feature & Rest wavelength [$\mu$m] & Line Center [$\mu$m] & Line Flux
[10$^{-22}$W cm$^{-2}$]  \\  \tableline
PAH & 7.7 & 15.9 & 14.5$\pm$4.4  \\
PAH & 8.6 & 17.4 & 12.4$\pm$1.8  \\
\null[\ion{S}{4}] & 10.5 & 21.7 & 1.2$\pm$1.8 \\
PAH & 11.3 & 23.1 & 46.1$\pm$5.3 \\
PAH & 12.7 & 26.0 & 3.0$\pm$0.6 \\ 
\null[\ion{Ne}{2}]\tablenotemark{a}  & 12.8 & 26.9 & 3.3$\pm$0.4 \\ 
\null[\ion{Ne}{5}]& 14.3 & 29.7 & 0.5$\pm$0.3  \\
\tableline
\end{tabular}
\\\smallskip
\tablenotetext{a}{The [\ion{Ne}{2}] emission line could be blended
with the 12.7\,$\mu$m PAH feature.} 
\label{tab:spectrum}
\end{center}
\end{table}

\end{document}